\documentstyle[12pt]{article}
\def\epm{$e^+e^-$}
\def\npe{$N_{pe}$}
\def\ufs{$\Upsilon (4{\rm S})$}
\input psfig
\begin{document}
\centerline
{\bf
THE RING IMAGING DETECTOR FOR CLEO III
}
\vskip 2pt
\centerline
{Marina Artuso}
\footnotesize
\centerline{\it Department of Physics, Syracuse University, Syracuse, NY 13244}
\normalsize
\vskip 3pt
\begin{center}
{\bf ABSTRACT}
\end{center}
\footnotesize
The CLEO detector will undergo major improvements in conjunction with a high
luminosity upgrade of the CESR electron positron collider, that should
increase the luminosity of this machine  by a factor of 10.
The most innovative feature of the planned CLEO detector is the addition of a
state of the art particle identification system, based on a proximity focused
Ring Imaging Cherenkov detector. The goal is to achieve good
hadron identification at all the momenta relevant to the study of decays of B
mesons produced at the \ufs\ resonance.
This detector design will be discussed, including details of
the mechanical design and the readout electronics.
The performance of a prototype module will be described.
\normalsize
\section*{\normalsize\bf 1. Introduction}
A major upgrade of the CLEO detector is underway (CLEO III), which, together
with the planned increase in luminosity of  the CESR \epm collider
at Cornell University, should open exciting prospects for  studying
CP violation in charged $B$ decays and mapping the  phenomenology
of rare $B$ decays. The main innovation  is the introduction of a high
class particle identification system which will distinguish charged
hadrons in the
kinematics domain characteristic of $B$ decays at the $\Upsilon
(4S)$  with high efficiency and low fake rate.  The
approach chosen is a Ring Imaging Cherenkov detector (RICH),  where the
position of the Cherenkov photons generated by relativistic particles
crossing a dense medium is reconstructed at a detector plane.

 The CLEO RICH is based on the proximity focusing approach, in  which the
Cherenkov cone is simply let to
expand in a volume filled with inert gas (the expansion gap) as much as allowed
by  other spatial constraints, before
intersecting the detector surface where the  coordinates of the Cherenkov
photons are reconstructed. In our approach  the radiator is a crystal (LiF).
 The
photon detection is performed in a thin gap  multiwire proportional
chamber (MWPC) with cathode pad readout. The photoconverter chosen is
triethylamine (TEA),
whose molecules are dispersed in the $CH_4$ gas of the photodetector.
A fine  segmentation of the cathode pad is required in order to achieve
the spatial  resolution needed, which in turn implies a high density of
readout electronics. Our design involves 230,000 cathode pads. The
availability of a suitable VLSI front end analog processor is therefore a key
element in this detector.

The goal of having excellent separation power
between  hadron from $B$  decays, most notably $\pi$'s and $K$'s from
2-body rare  decay modes,
determines the requirements on the angular resolution per
track. We define the separation in terms of the number of standard
deviations $N_{\sigma}$
defined as the ratio between the difference ${\theta _C(K)-\theta_C(\pi )}$
and the average error on the Cherenkov angle
measurement for each track, ${1/2[\sigma_{\theta_C}(K)+
 \sigma_{\theta_C}(\pi )]}$. Our design goal is $N_{\sigma}$ =4 for all the
momenta of interest.
 In our case the minimum angular separation between these two
particle species is 12.8 mr, corresponding to the maximum momentum of 2.8 GeV/c
at the \ufs . We consider
a mean number of 10 detected photoelectrons $N_{pe}$ as the minimum
acceptable value.
\section*{\normalsize\bf 2. A novel radiator geometry}
\vskip 2pt
In order to improve both the angular resolution per track and the number
of  reconstructed photoelectrons a novel radiator geometry has been
proposed [1], and is presently undergoing a technical feasibility
study. It involves  cutting the outer surface of the radiator
with a profile  resembling the teeth  of a saw, and therefore it is referred to
as
'sawtooth radiator'. The major advantage of this configuration is that it
reduces
the losses of photons due to total internal reflection at the interface between
the radiator and the expansion gap.
 A  detailed  simulation has
shown that a tooth angle of 45$^\circ$ is close to optimal.  Fig. 1 summarizes
the
results of this study, which focuses on particles with $p=2.8$ GeV/c.
The mean number of reconstructed photoelectrons \npe ,
Cherenkov angle resolution per track $\sigma _{\theta _c}$ and the probability
for a $\pi$ to fake a $K$ are plotted as a function of the angle $\theta$
between the charged particle and the normal to the radiator inner surface.
Besides featuring an improved performance, this solution has the
advantage of not requiring tilted radiator segments in the region
around $\theta =0$, in order to prevent all the Cherenkov photons from being
trapped
inside the radiator because of total internal reflection.
\section*{\normalsize\bf 3. Mechanical design}
The  mechanical design of this detector
faces several challenges.  One of the  most severe constrains is
dictated by the bandwidth of the photosensitive  element, centered around 150
nm. This implies a hermetic  sealing of the expansion gap from
the neighboring gas volumes, in order to  prevent contamination from
oxygen and water from the outside  environment. In addition the TEA inside the
photodetector must not leak into the
expansion gap because this would also cause loss of photoelectrons.
The other important goal is to keep the ${\rm  CaF}_2$ windows
free of any kind of mechanical stress in order to prevent  cracks from
developing. This has been achieved by attaching these windows to  their
frames through hinges, as shown in Fig. 2, which provide a soft joint to
relieve the stresses. The LiF radiators
are held in place by an inner carbon fiber cylinder to which they are
attached. Individual photosensitive detectors will be held in place by a
support
frame attached to the inner cylinder, as shown in Fig. 3.  In addition, the
back of the cathode boards which constitute
the outer side of the MWPC are   strengthened by hollow G10 rods which
also act as channels for the  cooling  gas ($N_2$).
The strength has been achieved with great care to minimize the amount of
material in the detector, in order to preserve the excellent performance of
the electromagnetic CsI calorimeter. The average material thickness is
13\% of a radiation length for tracks at normal incidence.

In order to improve the coupling between cathode
pads and anode wires and to reduce the mean number of pads
corresponding to a single photoelectron avalanche, the chamber geometry is
asymmetric, with the wire to cathode pad distance of 1 mm and an overall
gap thickness of 4 mm. The wires run along the 2.5 meter detector length.  In
order to preserve a high uniformity in the wire to cathode
distance over such a length ceramic spacers are glued between the wires and the
cathode pads along
the azimuthal direction every 30 cm.
\vskip 2pt
\section*{\normalsize\bf 4. Readout Electronics}
 The 230,000 readout channels are distributed over the
surface at the outer radius of the
detector and are impossible to access routinely. Therefore the
readout architecture must feature high parallelism, and
extensive testing of active components and
connection elements is required in order to insure their reliability.
 The MWPC detector surface is segmented into 30 modules, which will be
divided into 12 subunits each with 640 readout channels. Each of
these subunits will communicate via a low mass cable connection with VME
cards providing the control signals and receiving the analog or
digitized signals as discussed below.

Several considerations affect the design of the individual channel
processor. Low noise is an essential feature because the charge
probability distribution for the avalanche produced by a single photon
is exponential at moderate gains. It should be stressed that it
is beneficial to run at low gains to improve the stability
of chamber operation. Therefore in order to achieve
high efficiency, the noise threshold should be as low as
possible. An equivalent noise charge of about 200 electrons is adequate
for our purposes. On the other hand, an exponential distribution implies
 that a high dynamic range is desirable in order to preserve the
spatial resolution allowed by charge weighting.
Note that charged  particles are expected to  generate pulses at least 20 times
higher than the single photon mean pulse height.
In order to improve the robustness of the readout electronics against
sparking, a protection circuit constituted by a series resistor and two reverse
biased diodes is required in
 the input stage. Finally it is
 important to sparsify the information as soon as possible in the
processing chain, as the occupancy of this detector is very low and
therefore only a small fraction of the readout channels contain useful signals.

There is a preamplifier/shaper VLSI chip developed for
solid state detector applications which incorporates many of the
features discussed above [2]. A dedicated version of this chip,
called VA\_RICH, has been developed and will be tested shortly. Its predicted
equivalent noise
charge is given by:
\begin{equation}
ENC=\sqrt{(73 e^- + 12.1 e^-/pF)^2+50^2}
\end{equation}
The first term corresponds to the noise contribution from the
input transistor and the (80 $\Omega$) series resistor used for the input
protection
 and the second  to the noise from
subsequent stages, small but non negligible because of the lower gain
chosen to increase the dynamic range. This device is expected to maintain
linear response up to an input charge of 700,000 $e^-$.

The choice of digitization and sparsification technique has not yet been
finalized. Under consideration is the digitization and sparsification at the
front end level, using the zero suppression scheme and the ADC included
in the SVX II readout chip [3]. Alternatively we will digitize all the analog
output signal and perform the zero suppression afterwards.
\section*{\normalsize\bf 5. Performance of the CLEO RICH Prototype}
Our first step in the R\&D effort towards the construction of this
detector has been the construction of a  prototype whose
length is about 1/3 of an individual detector module in the final design
and whose width is about the same. The prototype system is enclosed in a leak
tight
aluminum box. The
expansion gap is 15.7 cm. In this prototype we use plane LiF radiators.
 The chamber geometry is approximately the same as the final design
in terms of gap size, wire to cathodes distance and pad sizes. The total
number of pads read out is 2014. Pad signals are processed
by VA2 [4] preamplifier and shapers. The detector plane is divided
into 4 quadrants each of which has 8 VA2 daisy--chained for serial readout.

This prototype has been installed in a cosmic ray set up
composed of a telescope of scintillators whose geometrical
arrangement can be varied. We either trigger on
energetic cosmic rays having their Cherenkov image within the acceptance
of the photodetector but not the charged track, or we trigger on charged tracks
going through the detector.

Fig. 4 shows the charge distribution of reconstructed photon clusters as
a function of the anode wire voltage $V_a$ for several different voltages
when the gas mixture is $CH_4-TEA$.
Note that the pulse height distribution is consistent with an exponential
profile and its mean value increases with $V_a$, as expected. Fig. 5
shows the excitation curve measured for the same gas mixture. It can be
seen that the plateau corresponds to $N_{pe}\approx 13$. This number has
to be corrected for possible background hits, which we estimate to be
about 1 per event. On the other hand there are subtle effects related to
the fact that clusters associated with different photoelectrons are
often close in space and therefore have a finite chance of
overlapping  into a single cluster, thus reducing the reconstructed \npe .
So far we believe that we are loosing on average one
photoelectron per event due to this effect and that some improvement can
be achieved in the future. This performance is in close agreement with
our expectations based on the performance of a similar prototype built
and tested by the College de France--Strasbourg group [5]. A
tracking system is being added to this set--up to allow us to measure
also the angular resolution of our device.
\section*{\normalsize\bf 6. Acknowlegements}
We would like to thank the
A. Efimov, M. Gao, S. Kopp, R. Mountain, Y. Mukhin, S. Playfer, and S. Stone
for
for useful discussion. Special thanks are
due to E. Nygard for low noise electronics and all the team of IDE AS for
developing the VA\_RICH chip optimized
for our application.

{\bf REFERENCES}
\begin{description}
\footnotesize
\item{[1]} {\sc A. Efimov} {\it et al.}, {\it Syracuse Un. Preprint},
HEPSY-94-8 (1994) to be published in {\it Nucl. Instr. and Meth. A}
\item{[2]} {\sc E. Nygard} {\it et al.}, {\it Nucl. Instr. and Meth. A301}
(1991) 506.
\item{[3]} {\sc O. Milgrome} {\it Talk given at the 2nd International Meeting
on Front End Electronics for Tracking Detectors at Future High Luminosity
Colliders} Perugia, Italy (1994).
\item{[4]} {\sc O. Toker} {\it et al.}, {\it CERN Preprint CERN--PPE/93--141}
(1993).
\item{[5]} {\sc J.L. Guyonnet} {\it et al.}, {\it Nucl. Instr. and Meth. A343}
(1994) 178.
\end{description}
\eject
\normalsize
\centerline{Figure Captions}
Fig. 1 Performance of sawtooth radiator (curves are parametrized by the
sawtooth angle):
1.a) The average number of detected photoelectrons as a function of the
incident track angle, 1.b)The
angular resolution per track as a function of incident track angle, 1.c) The
probability for a
2.8 GeV/c $\pi$ to fake a $K$ for 95\% $\pi$ efficiency.

Fig. 2  A detail of a section of the CLEO RICH showing the mechanical
design of individual detector modules showing the attachment method for
the $CaF_2$ window and the method to strengthen the cathode plane.

Fig. 3 A detail of a section of the CLEO RICH showing the attachment of the
support structure
for the detector modules to the inner cylinder.

Fig.4: Photon induced avalanche charge distribution at different anode
voltages. The voltage on the metallization of the $CaF_2$ windows is
kept at -1350V.

Fig. 5: Excitation curve for $CH_4-TEA$. The voltage on the metallization
of the $CaF_2$ windows is kept at -1350V.
\eject
\begin{figure} [htbp]
\vspace{-4cm}
\centerline{\psfig{figure=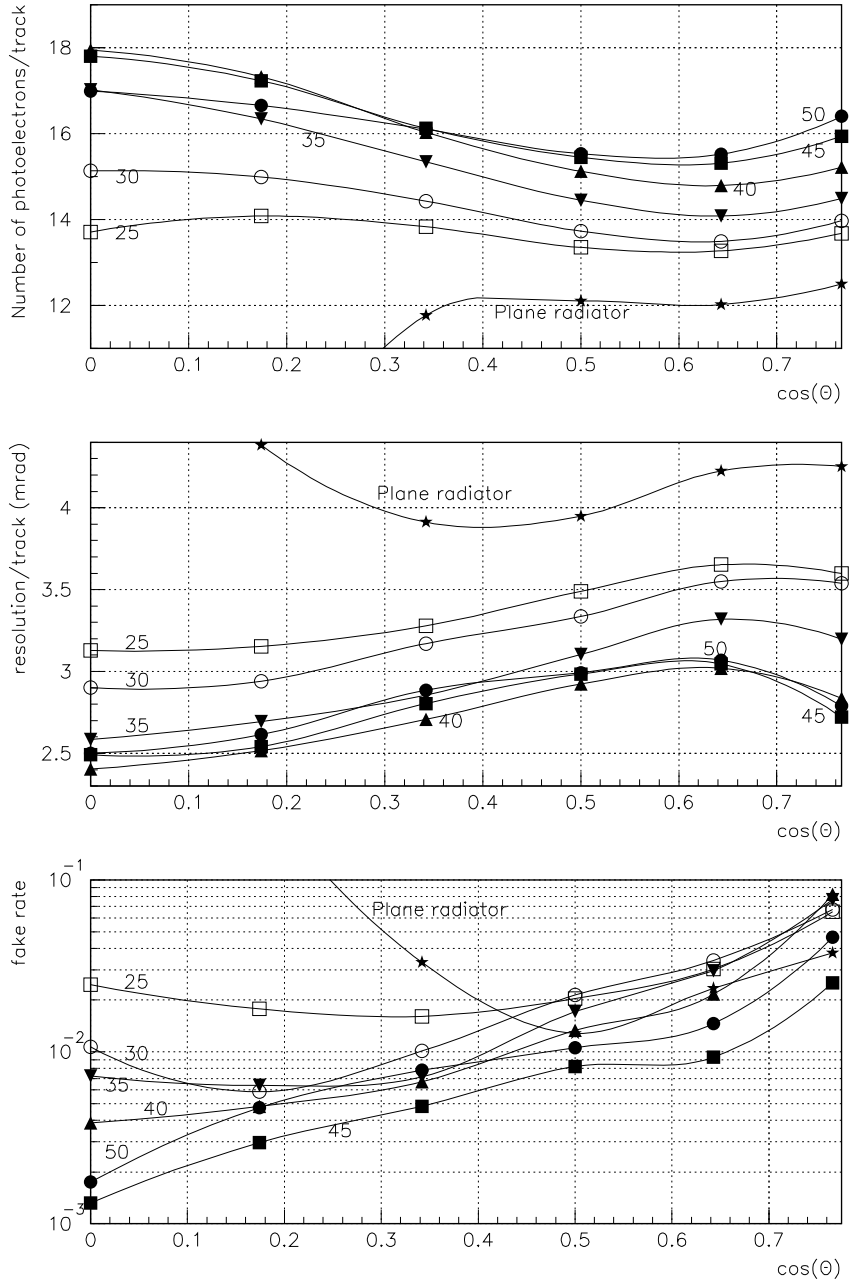,height=7.5in,bbllx=0bp,%
bblly=0bp,bburx=600bp,bbury=800bp,clip=}}
\caption{Performance of sawtooth radiator (curves are parametrized by the
sawtooth angle):
1.a) The average number of detected photoelectrons as a function of the
incident track angle, 1.b)The
angular resolution per track as a function of incident track angle, 1.c) The
probability for a
2.8 GeV/c $\pi$ to fake a $K$ for 95\% $\pi$ efficiency.}
\end{figure}
\clearpage
\begin{figure}[hbt]
\vspace{1.0cm}
\centerline{\psfig{figure=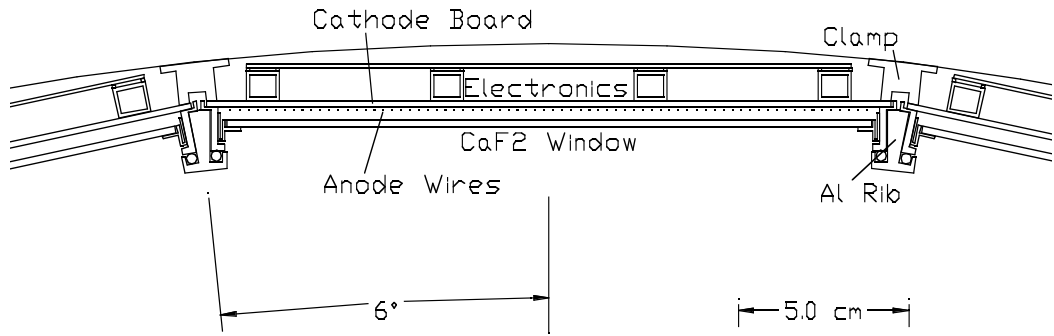,height=7in,bbllx=0bp,%
bblly=0bp,bburx=600bp,bbury=700bp,clip=}}
\vspace{-6.0cm}
\caption{ A detail of a section of the CLEO RICH showing the mechanical
design of individual detector modules showing the attachment method for
the $CaF_2$ window and the method to strengthen the cathode plane.}
\end{figure}
\begin{figure}[hbt]
\centerline{\psfig{figure=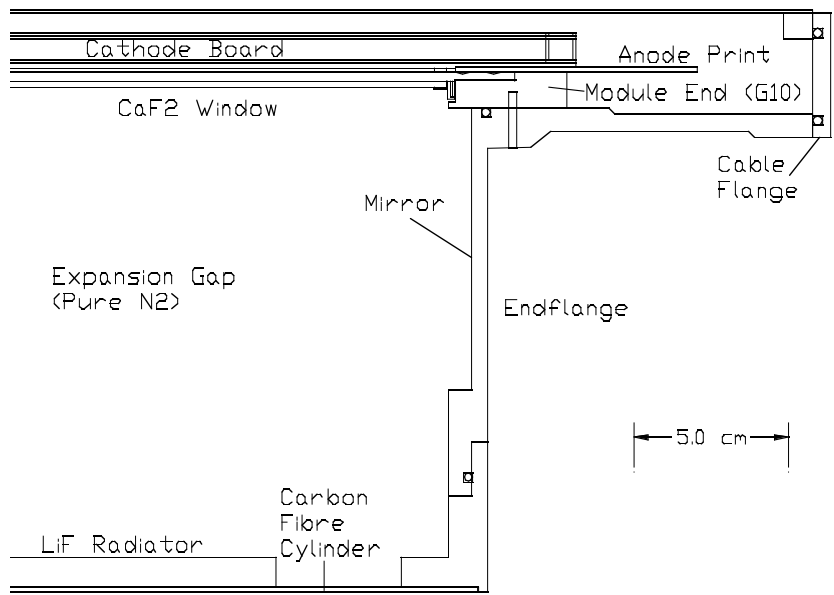,height=7in,bbllx=0bp,%
bblly=0bp,bburx=600bp,bbury=800bp,clip=}}
\vspace{-6.0cm}
\caption{ A detail of a section of the CLEO RICH showing the attachment of the
support structure
for the detector modules to the inner cylinder.}
\end{figure}
\clearpage
\begin{figure}
\centerline{\psfig{figure=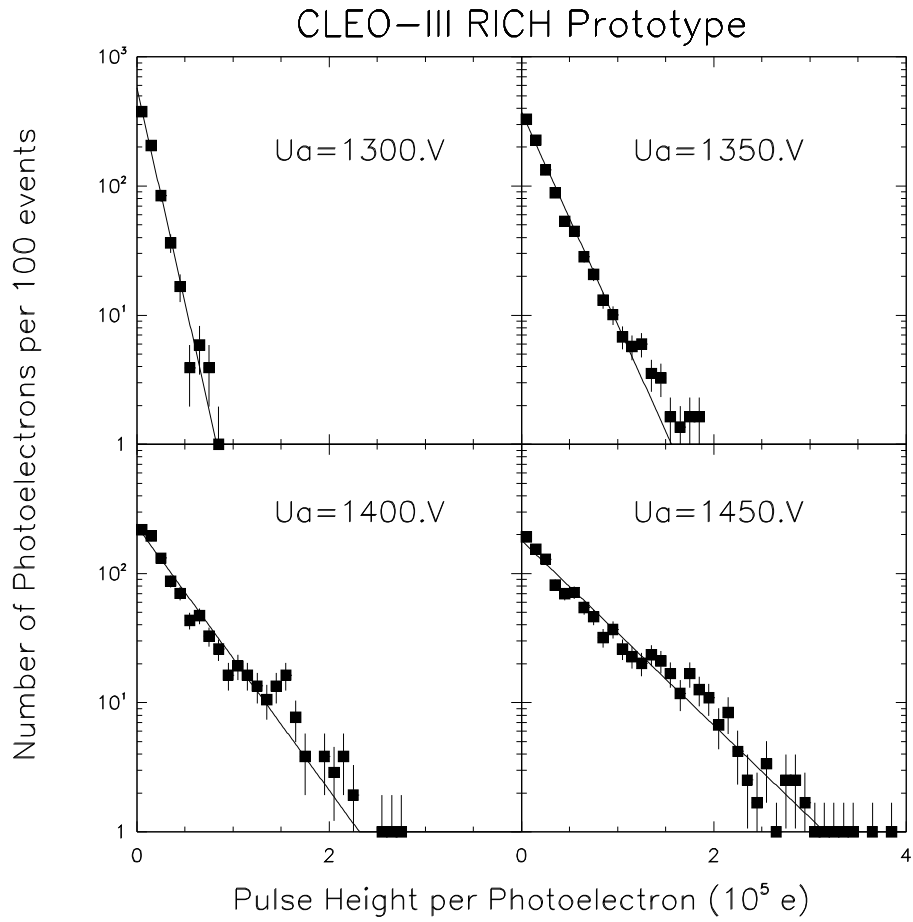,height=7in,bbllx=0bp,%
bblly=0bp,bburx=600bp,bbury=700bp,clip=}}
\vspace{-1.0cm}
\caption{ Photon induced avalanche charge distribution at different anode
voltages. The voltage on the metallization of the $CaF_2$ windows is
kept at -1350V.}
\end{figure}
\eject
\begin{figure}
\vspace{-6cm}
\centerline{\psfig{figure=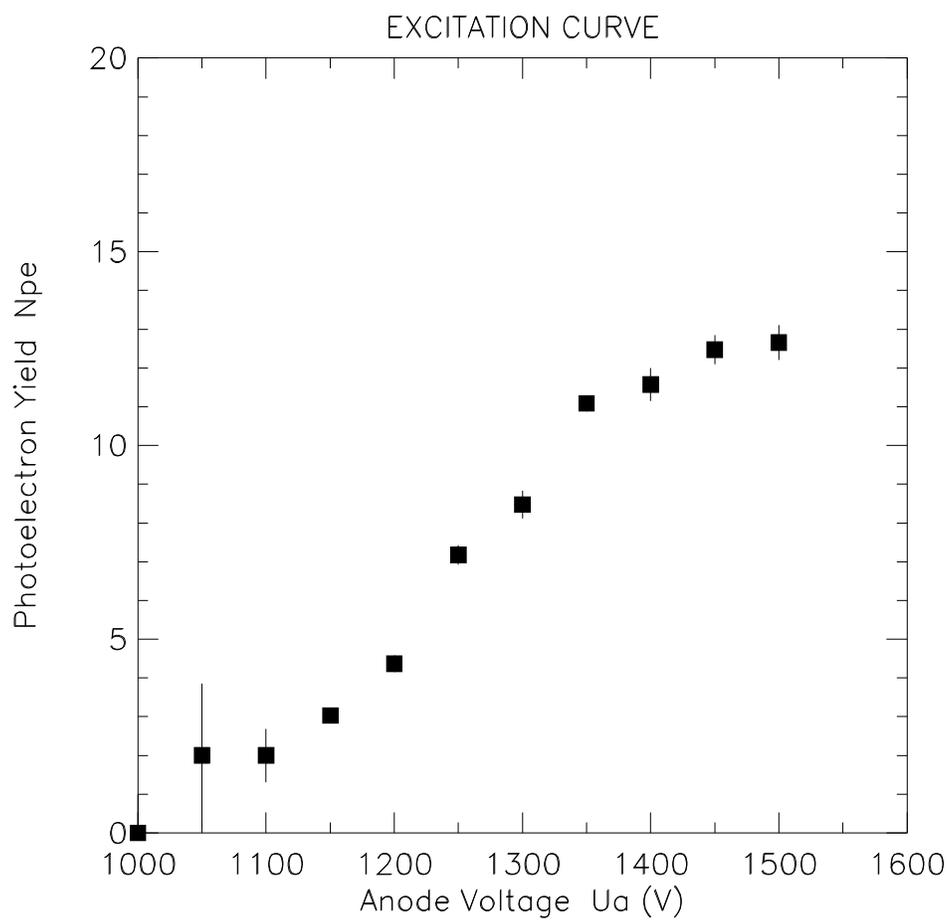,height=7in,bbllx=0bp,%
bblly=0bp,bburx=600bp,bbury=700bp,clip=}}
\caption{ Excitation curve for $CH_4-TEA$. The voltage on the metallization
of the $CaF_2$ windows is kept at -1350V.}
\end{figure}
\eject
\end{document}